\documentclass[reprint,amsmath,amssymb,aps,prb,floatfix,showpacs]{revtex4-1}
\usepackage{graphicx}
\usepackage{dcolumn}
\usepackage{bm}
\usepackage{multirow}
\usepackage{color}
\usepackage[english]{babel}
\usepackage{amsmath}
\usepackage{amsfonts}
\usepackage{amssymb}
\usepackage{color}
\usepackage{tabularx,booktabs}
\newcolumntype{Y}{>{\centering\arraybackslash}X}
\usepackage{dcolumn}
\begin{document}

\title{Diffusive nature of thermal transport in stanene}
\author{Arun S. Nissimagoudar$^{1}$, Aaditya Manjanath$^{1,2}$ and Abhishek K. Singh$^{1,\ast}$}
\affiliation{$^1$Materials Research Centre, Indian Institute of Science, Bangalore - 560012, India}
\affiliation{$^2$Centre for Nano Science and Engineering, Indian Institute of Science, Bangalore - 560012, India}
\date{\today}

\begin{abstract}
Using the phonon Boltzmann transport formalism and density functional theory based calculations, we show that stanene has a low thermal conductivity. For a sample size of 1$\times$1 $\mu$m$^{2}$ ($L\times W$), the lattice thermal conductivities along the zigzag and armchair directions are 10.83 W/m-K and 9.2 W/m-K respectively, at room temperature, indicating anisotropy in the thermal transport. The low values of thermal conductivity are due to large anharmonicity in the crystal resulting in high Gr\"{u}neisen parameters, and low group velocities. The room temperature effective phonon mean free path is found to be around 17 nm indicating that the thermal transport in stanene is completely diffusive in nature. Furthermore, our study brings out the relative importance of the contributing phonon branches and reveals that, at very low temperatures, the contribution to lattice thermal conductivity comes from the flexural acoustic (ZA) branch and at higher temperatures it is dominated by the longitudinal acoustic (LA) branch. We also show that lattice thermal conductivity of stanene can further be reduced by tuning the sample size and creating rough surfaces at the edges. Such tunability in the lattice thermal conductivity in stanene suggests its applications in thermoelectric devices. 
\end{abstract}

\maketitle


\section{Introduction}
Two-dimensional (2D) group IV materials, such as graphene, silicene, and germanene, have attracted great attention, because of their novel properties~\cite{neto2009electronic,sarma2011electronic,balandin2008superior,balendhran2015elemental,
mas20112d,xu2014thermal} and potential application in nanoelctronic devices~\cite{neto2009electronic,balandin2011thermal,chen2013substrate,cai2010thermal,xu2014thermal,schwierz2015two}. Recently, a new 2D material from the same group, called stanene (monolayer of tin atoms), has garnered interest due to outstanding properties such as topological superconductivity~\cite{wang2014two} and room temperature quantum anomalous Hall (QAH) effect~\cite{wu2014prediction}. In addition, stanene was found to be a topological insulator with large-gap 2D quantum spin Hall (QSH) states at room temperature~\cite{liu2011} that enable superior electric conduction despite the spin orbit coupling (SOC) induced band gap of $\sim$0.1 eV~\cite{xu2013large}. These unique properties make stanene a promising material for application in next generation nanoelectronic devices. However, any possible application of a new material requires proof of its existence in terms of the synthesis as well as stability. The low-buckled form of stanene has been shown to be stable~\cite{sil-ger-buck-latt,cai2015,van2014} and was synthesized as early as last year, on the Bi$_\mathrm{2}$Te$_\mathrm{3}$ (111) substrate~\cite{zhu2015}. This recent synthesis of stanene could stimulate further experimental research into the novel electronic properties exhibited by it.

Together with the understanding of electronic properties, it is also very important to have a comprehensive knowledge of the thermal transport in a material, before it can be used in a given application. With the advent of novel fabrication techniques of nanoelectronic devices, heat removal has been a biggest challenge due to increased levels of dissipated power and speed of operation of the in-built circuitry. Materials with high thermal conductivity are ideal for such applications. On the other hand, a low value of thermal conductivity would make a material promising for waste energy harvesting via thermoelectric effect. Recent theoretical studies have shown that in addition to a high electrical conductivity~\cite{vandenberghe2014calculation}, stanene's thermoelectric figure of merit (ZT) can be strongly tuned by the sample size~\cite{xu2014enhanced}. Hence, an understanding of the thermal transport in stanene will determine its application and role in electronic or thermoelectric devices. Furthermore, due to the similarities and differences in the lattice structure between stanene and other existing group IV 2D materials (graphene, silicene, and germanene), it is of fundamental interest to investigate and find differences in the mechanism of thermal transport in these elemental group IV sheets.  

In this work, we investigate the lattice thermal conductivity of stanene by combining the first-principles calculations and phonon Boltzmann transport formalism under relaxation time approximation. We take into account, the anharmonic three-phonon (normal as well as umklapp), and the angle-dependent phonon-boundary scattering mechanisms. The inclusion of the angle-dependence in our formalism allows us to explore the thermal transport along zigzag, ZZ ($\theta=$~0$^\circ$) and armchair, AC ($\theta=$~30$^\circ$), directions. Based on this formalism, we show that the lattice thermal conductivity of stanene is low owing to very high Gr\"{u}neisen parameter and very low group velocity in this material. At room temperature, the maximum contribution to lattice thermal conductivity comes from longitudinal acoustic (LA) branch (91$\%$), unlike graphene, where the flexural acoustic (ZA) branch is the main contributor. Furthermore, phonon thermal conductivity of stanene can be further reduced by tuning the sample size and creating rough surfaces at the edges. Therefore, the performance of the stanene based thermoelectric devices can be significantly enhanced by this unprecedented tunability of thermal conductivity.

\section{Methodology} 

\subsection{Lattice thermal conductivity}
According to Fourier's law~\cite{klemens1958solid}, the lattice thermal conductivity, $\kappa_{\alpha \beta}$, of a material is defined by:
\begin{equation}
U_\alpha = -\kappa_{\alpha \beta} \nabla T_\beta
\label{eq:1}
\end{equation}
where $U_\alpha$ is the heat flux density along the $\alpha^{\text{th}}$ direction produced by the temperature gradient $\nabla T_\beta$ along the $\beta^{\text{th}}$ direction. Considering heat to be transported through a stanene sheet of length $L$ and width $W$, $\kappa_{\alpha\beta}$ can be obtained by solving the phonon Boltzmann transport equation under the relaxation time approximation and is given by~\cite{bandari1988,callaway1959model,srivastava1990physics}:
\begin{equation}
\kappa_{\alpha \beta}=\frac{1}{LW\delta}\sum_{s,\mathbf{q}} \hbar\omega_{s}(\mathbf{q}) v_{g\alpha}(s,\mathbf{q})\\
v_{g\beta}(s,\mathbf{q})\tau_{s}^{C}(\mathbf{q})\dfrac{\partial \bar{n}}{\partial T}
\label{eq:2}
\end{equation}

where $\delta$ is the thickness of the layer and is chosen as 4.5~\AA, equal to the van der Waals diameter of tin atoms, $\omega_{s}(\mathbf{q})$ is the phonon frequency corresponding to a branch $s$, $v_{g\alpha}(s,\mathbf{q})$ and $v_{g\beta}(s,\mathbf{q})$ are the components of phonon group velocity along the $\alpha$ and $\beta$ directions, $\tau_{s}^{C}$ is the effective relaxation time that incorporates the momentum conserving (normal), non conserving (umklapp) three-phonon processes and sample dependent scattering processes, and $\bar{n}$ is the equilibrium phonon population given by the Bose-Einstein distribution. Along a particular chiral direction, \textit{j}, the expression for $\kappa$ is obtained by  converting the summation to the integral, and taking into account the two-dimensional phonon density of states, given by~\cite{chen2005nanoscale}:

\begin{multline}
\kappa_{j}=\frac{\hbar^{2}}{2k_{B}T^{2}\pi\delta}\sum_{s} \int_{0}^{\omega_{max}} \int_{0}^{2\pi} d\omega d\theta \omega_{s}^{2} \lbrace v_{g}(\omega_{s})\rbrace_{j}^{2}\\
\times \tau_{s}^{C}(\omega)\bar{n}(\bar{n}+1)g(\omega_{s})
\label{eq:3}
\end{multline}

Here, $\omega_\text{max}$ is the maximum cut-off frequency, $\theta$ is the angle of the motion of the phonons relative to the main thermal transport direction, $\lbrace v_{g}(\omega_s)\rbrace_{j}$ is the component of the phonon group velocity in the $\textit{j}^\text{th}$ direction given by $\lbrace v_{g}(\omega_s)\rbrace_{j}=\left|\frac{d\omega_{s}}{dq_{j}}\right|\cos(\theta)$, and $g(\omega_{s})$ is the total phonon density of states.

\subsection{Relaxation rates}
The major aspect in the calculation of lattice thermal conductivity is the relaxation time, $\tau_{s}^{c}(\omega)$ arising from different phonon scattering mechanisms. The phonons may undergo collisions with sample boundaries, defects, impurities and other phonons. Different scattering mechanisms may dominate at different temperatures~\cite{alofi2013thermal}. In the temperature range of interest (2~$< T < $~1000 K), the resistive scattering processes limiting thermal conduction in isotopically pure stanene include the scattering of phonons by sample boundaries and other phonons (normal and Umklapp processes). While the phonon-phonon scattering mechanisms are characteristic of the material, the phonon-boundary scattering mechanisms are sample-dependent. 

\subsubsection{Phonon-phonon scattering}
The phonon-phonon interactions become dominant at high temperatures and play an important role in limiting the lattice thermal conductivity. The phonon-phonon interactions contributing to the thermal resistance in a crystalline material are the normal and Umklapp three-phonon processes~\cite{klemens1958solid,alofi2013thermal}. The three-phonon process relaxation time depends on the nature of the phonon spectrum~\cite{bandari1988,morelli2002estimation}. However, the normal phonon processes do not contribute directly to the thermal resistance but are crucial in spreading out the influence of the other resistive processes to the entire phonon spectrum. In the long wavelength approximation, these anharmonic three phonon-phonon relaxation times can be expressed by phenomenological relationships and are known to explain the experimental data very well for bulk and low-dimensional semiconductors ~\cite{morelli2002estimation,slack1964thermal,nika2009lattice,sadeghi2012thermal,liu2014anomalous}. These expressions are given by~\cite{morelli2002estimation,slack1964thermal}:
\begin{equation}
(\tau_{s}^{N}(\omega_{s}))^{-1} = B_{sN}\omega_{s}^{2}T^{3}
\label{eq:4}
\end{equation}
and
\begin{equation}
(\tau_{s}^{U}(\omega_{s}))^{-1} = B_{sU}\omega_{s}^{2}T\exp(-\Theta_{Ds}/3T)
\label{eq:5}
\end{equation}
The scattering rate coefficients $B_{sN}$ and $B_{sU}$ correspond, respectively, to the phonon-phonon normal and umklapp scatterings, in the material which are defined by~\cite{morelli2002estimation}: $B_{sN} = (k_{B}/\hbar)^{3}[\hbar \gamma_{s}^{2}V/M((|d\omega_{s}/dq_{j}|)^{5}]$ and $B_{sU} =\hbar\gamma_{s}^{2}/M(|d\omega_{s}/dq_{j}|)^{2}\Theta_{Ds}$, where $M$ is the mass of the unit cell, $V$ is its volume, $\gamma_{s}(\mathbf{q})=-\frac{V}{\omega_{s}(\mathbf{q})}\frac{\partial\omega_{s}(\mathbf{q})}{\partial V}$ is the Gr\"{u}neisen parameter for phonons of a particular branch $s$ derived from the phonon dispersion relations of the material system, and $\Theta_{Ds}$ is the Debye temperature, which is obtained from the maximum cutoff frequency for each branch.

\subsubsection{Phonon-boundary scattering}
The scattering of phonons by the boundaries which becomes dominant at low temperatures occur at the two terminal edges (along the transport direction) and two lateral edges (perpendicular to the transport direction) of the sample. The expressions for the relaxation rates are taken as ~\cite{bae2013ballistic,aksamija2012thermal,liu2015anisotropic,liu2014anomalous}
\begin{equation}
(\tau_{s}^{L}(\omega_{s}))^{-1} = \frac{|d\omega_{s}/dq_{j}|\cos(\theta)}{L}
\label{eq:6}
\end{equation}
and
\begin{equation}
(\tau_{s}^{W}(\omega_{s}))^{-1} = \frac{|d\omega_{s}/dq_{j}|\sin(\theta)}{W}\frac{1-p}{1+p}
\label{eq:7}
\end{equation}    

These expressions are based on the partially specular interaction of phonons boundaries~\cite{srivastava1990physics,bae2013ballistic}. Depending on the nature of the boundary, the phonon can either get reflected resulting in specular interaction which only flips the phonon momentum about the boundaries without randomizing it, or get scattered diffusively by randomizing the phonon momentum~\cite{bae2013ballistic}. The fraction of phonon interaction with boundaries that is specular is given by the specularity parameter, $p$. The specularity parameter gives the probability of scattering and its value varies from 0 to 1, with $p=$~0 representing a completely rough surface leading to completely diffusive scattering and $p>$~0 where partial specular reflection occurs, thereby decreasing the effect of the boundary scattering. In eqn~\ref{eq:6}, we have assumed the terminals to be extremely rough ($p=$~0) where phonon-boundary scattering is completely diffusive. On the other hand, in eqn~\ref{eq:7} we have assumed partial diffusive scattering with $p$ ranging between 0 and 1. 

\subsection{Details of first-principles calculations}
In order to calculate $\kappa$ accurately, a detailed knowledge of the full-band phonon dispersion ($\omega_s$) and phonon group velocities ($v_g(\omega_s)$) is necessary. These are obtained by first-principles calculations within the density functional theory (DFT) and small displacement method (SDM). DFT calculations are performed by employing the Vienna \textit{ab initio} simulation package (VASP)~\cite{kresse1996efficiency,kresse1996efficient}. The exchange and correlation part of the total energy is approximated by the generalized gradient approximation (GGA) using the Perdew-Burke-Ernzerhof (PBE)~\cite{perdew1996generalized} type of functional. The optimized structure of low-buckled stanene with a primitive cell configuration of two atoms is shown in Fig.~\ref{Fig:1}(a), which is geometrically similar to silicene~\cite{sil-ger-buck-latt,meng2013buckled}, germanene~\cite{sil-ger-buck-latt,derivaz2015continuous}, and blue phosphorene~\cite{blue-phos}. The obtained lattice constant of 4.67~\AA~is in good agreement with previous studies~\cite{sil-ger-buck-latt,xu2013large}. A vacuum of 15~\AA~is used along the $z$-axis to avoid spurious interactions between the sheet and its periodic images. Structural relaxation was performed using the conjugate-gradient method until the absolute values of the Hellman-Feynman forces were converged to within 0.005 eV\AA$^{-1}$. Integrations over the Brillouin zone were performed using a well-converged Monkhorst-Pack~\cite{MP_kpoint} \textbf{k}-mesh of 15$\times$15$\times$1. Phonon dispersion calculations were performed using SDM, with a supercell size of 5$\times$5$\times$1, an energy cut-off of 450 eV and an energy tolerance value of 1$\times$10$^{-8}$ eV in order to obtain accurate forces. The Harmonic interatomic force constants (IFCs) were extracted from the SDM calculation using the Phonopy package~\cite{phonopy}.
 
\section{Results and discussion}
\subsection{Phonon dispersions}
\begin{figure}[!ht]
\includegraphics[width=\columnwidth]{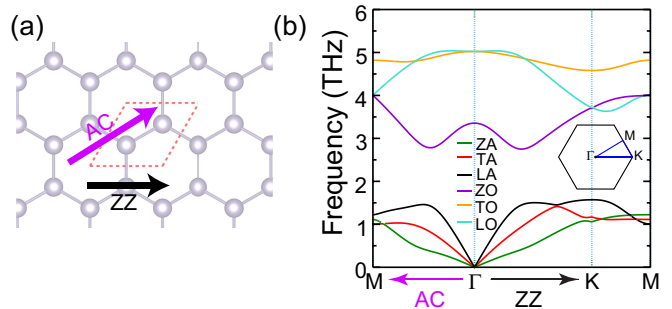}
\caption{(a) Top view of stanene. The unit cell is indicated by the dashed red line. The directions of thermal transport (zigzag and armchair) are also shown. (b) Full band phonon dispersion for stanene along high-symmetry directions as shown by the blue lines in the Brillouin zone (inset).}
\label{Fig:1}
\end{figure}
The phonon dispersion is calculated along the high symmetry directions, as shown in Fig.~\ref{Fig:1}(b). The directions $\Gamma$-M and $\Gamma$-K in the reciprocal space correspond to the AC and ZZ directions in the real space. Due to two atoms in the primitive unit cell of stanene, there are six phonon branches, out of which three are acoustic phonon branches-longitudinal acoustic (LA) and transverse acoustic (TA) branches in the basal plane, and flexural acoustic (ZA) branch, perpendicular to the basal plane-which are separated by a gap of around 1.25 THz below non-polar optical branches-longitudinal optic (LO), transverse optic (TO), and flexural optic (ZO) branches. In stanene, all acoustic branches have a linear dependence in $\mathbf{q}$ near the $\Gamma$ point. However, other 2D materials such as graphene~\cite{nika2009phonon,bae2013ballistic}  show a purely quadratic dependence while silicene~\cite{gu2015first} and blue phosphorene~\cite{jain2015strongly} do not display a purely quadratic dependence in $\mathbf{q}$ for the ZA branch but linear dependence for the LA and TA branches. This behavior in blue phosphorene, silicene and stanene, is due to the $sp^\text{2}$-$sp^\text{3}$ hybridization, which results in a periodic up and down motion of atoms in the direction perpendicular to the basal plane ($\pm z$) forming a trigonal arrangement of atoms belonging to $D_{3d}$ point group~\cite{ribeiro2015group}. This corresponds to a threefold rotational symmetry which breaks the reflection symmetry of atomic vibrations over the basal plane. Therefore, the ZA branch not only contains out-of-plane motions of atoms, but also includes components of in-plane motions, which leads to a non-quadratic phonon dispersion near the $\Gamma$ point. To understand such behavior we have calculated the phonon dispersions for graphene, blue phosphorene, and silicene (Fig. S1 in ESI$\dag$) using SDM and analyzed the force constants for each elemental sheet. Numerical fitting for the ZA branch in the long wavelength limit shows approximate relations as $\omega \sim |\mathbf{q}|^{2}$, $\omega \sim |\mathbf{q}|^{1.8}$, $\omega \sim |\mathbf{q}|^{1.56}$ and, $\omega \sim |\mathbf{q}|^{1.1}$, respectively for graphene, blue phosphorene, silicene and stanene. The non-quadratic relations of frequency clearly indicate that the ZA branch contains in-plane components. Analysis of force constants of blue phosphorene, silicene and stanene show that the off-diagonal ($xz$) components are non-zero, while it is zero for graphene. On comparing the magnitudes of the off-diagonal components ($xz$) of the force constants for these sheets, we find that stanene has a much larger value suggesting that the out-of-plane motions of Sn atoms are more strongly coupled with motions along the $x$ direction. As a result, this leads to a linear dispersion in the long wavelength limit. We further tried to understand this phenomenon by studying the nature of bonding between atoms through the electron localization function (ELF) for these sheets (Fig. S2 in ESI$\dag$). The ELF represents the probability of finding a pair of electrons at a given position. The values range between 0 and 1, which denotes no (0) to complete localization (1). As shown in Fig. S2(a)(ESI$\dag$), for graphene, the electrons are localized at the center of a C-C bond implying a strong covalent bond formation ($sp^\text{2}$ hybridization). However, in the case of blue phosphorene and silicene, this localization is more diffused (Figs. S2(b) and (c) in ESI$\dag$) indicating a lesser covalency, therefore, a dehybridization of purely $sp^\text{2}$ bonding into $sp^\text{3}$-like nature. This enables a coupling of the out-of-plane motion with the in-plane motion of atoms. For stanene, the electrons are more delocalized compared to the other cases leading to increased tendency towards $sp^\text{3}$ hybridization and a stronger coupling between the out-of-plane motion and in-plane motion of atoms. This leads to the linear dependence of the ZA branch in the long wavelength limit for stanene.              

\subsection{Phonon group velocities and Gr\"{u}neisen parameters}
\begin{figure}[!t]
\includegraphics[width=\columnwidth]{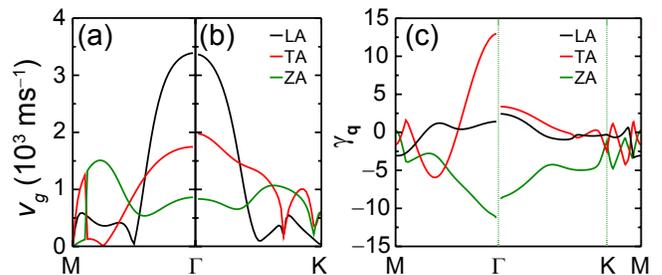}
\caption{Acoustic phonon group velocities along (a) armchair ($\Gamma$-M) (b) zigzag ($\Gamma$-K) directions. (c) Full band Gr\"{u}neisen parameters of stanene along the high-symmetry directions.}
\label{Fig:2}
\end{figure}

Having obtained the phonon frequencies at each $\mathbf{q}$-point corresponding to each branch $s$, we calculate the phonon group velocities: $\mathbf{v}_g(s,\mathbf{q})=\frac{\partial\omega_s(\mathbf{q})}{\partial\mathbf{q}}$. Fig.~\ref{Fig:2}(a and b) show the phonon group velocities for LA, TA, and ZA branches along the AC and ZZ directions, respectively. While, for LA and ZA branches, the group velocities near the $\Gamma$ point along the AC direction (3,388 ms$^{-\text{1}}$ and 868 ms$^{-\text{1}}$) are not significantly different from those along ZZ (3,366 ms$^{-\text{1}}$ and 833 ms$^{-\text{1}}$), the group velocity for TA branch along AC is lower (1,745 ms$^{-\text{1}}$) compared to ZZ (1,977 ms$^{-\text{1}}$) indicating an anisotropy in this branch. On comparing the group velocities of stanene with those of other group IV elements, we find stanene to have low group LA and TA velocities (graphene; $v_\mathrm{LA}=$~21,400 ms$^\mathrm{-1}$ and $v_\mathrm{TA}=$~14,200 ms$^\mathrm{-1}$, silicene; $v_\mathrm{LA}=$~6,400 ms$^\mathrm{-1}$ and $v_\mathrm{TA}=$~3,700 ms$^\mathrm{-1}$, germanene; $v_\mathrm{LA}=$~4,031 ms$^\mathrm{-1}$ and $v_\mathrm{TA}=$~2,124 ms$^\mathrm{-1}$)~\cite{alofi2013thermal,hu2013anomalous}. Such low values are expected to lower the thermal conductivity of stanene.

In addition to phonon group velocities, the Gr\"{u}neisen parameter plays an important role in determining $\kappa$ at higher temperatures. It provides the required information on the anharmonic interactions between lattice waves and the extent of the phonon scattering. The Gr\"{u}neisen parameter of a material is strongly dependent on phonon polarization and wavevector. We evaluate it along the high symmetry directions (Fig.~\ref{Fig:2}(c)) by dilating the crystal volume by 0.25\% biaxial strain.
The $\gamma_{s}(\mathbf{q})$ value for the LA branch is found to be the lowest, followed by the ZA and TA branches. This leads to a high thermal conductivity for the LA branch. In addition, a higher value of $\gamma_{s}(\mathbf{q})$ along the AC direction in the case of the TA branch, compared to that along ZZ, suggests the dominance of phonon-phonon scattering mechanisms, leading to a suppression in the TA thermal conductivity along AC. Overall, the anisotropy in the Gr\"{u}neisen parameters follows the group velocity trends (Fig.~\ref{Fig:2}). The obtained Gr\"{u}neisen parameters of the LA and TA branches are the highest among the group IV elements (graphene; $\gamma_\mathrm{LA}$~=1.1 and $\gamma_\mathrm{TA}=$~0.7, silicene; $\gamma_\mathrm{LA}$~=1.2 and $\gamma_\mathrm{TA}=$~0.7, germanene; $\gamma_\mathrm{LA}$~=2.1 and $\gamma_\mathrm{TA}=$~3.1)~\cite{kong2009first,huang2015phonon} thereby, rendering the stanene sheet a low thermal conductivity material following the relationship $\kappa\propto 1/\gamma_s^2$.

\subsection{Lattice thermal conductivity}
Using the obtained values of the phonon frequencies and group velocities, we evaluate the lattice thermal conductivity ($\kappa$). In our numerical estimation of $\kappa$, we consider the phonons to be scattered by the boundaries of stanene and other phonons via normal and umklapp processes. In the evaluation of phonon-phonon relaxation rates (eqn~\ref{eq:4} and~\ref{eq:5}), the $\mathbf{q}$-dependence of the Gr\"{u}neisen parameters for each of the phonon branches are taken into account.

\begin{figure}[!t]
\includegraphics[width=\columnwidth]{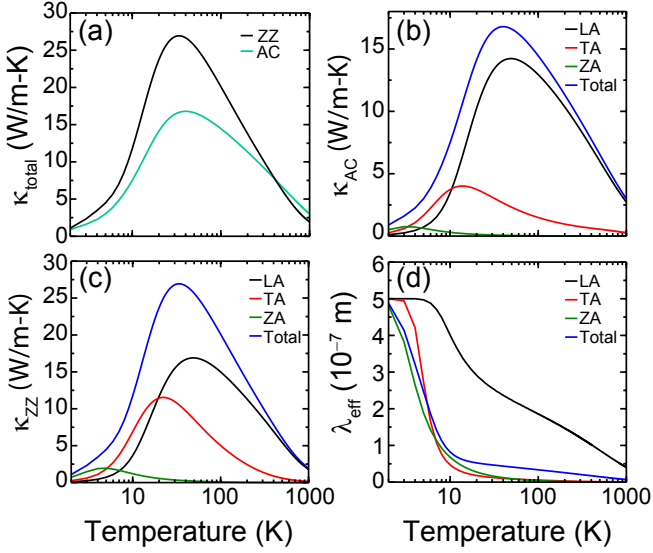}
\caption{Temperature dependence of (a) total lattice thermal conductivity ($\kappa_\text{total}$) of stanene along ZZ and AC directions, contributions to thermal conductivity from different phonon branches along (b) AC and (c) ZZ directions, and (d) effective mean free path along ZZ direction. All calculations have been performed assuming the following dimensions of the stanene sample$-$width ($W$): 1 $\mu$m and length ($L$): 1 $\mu$m. A specularity parameter of $p=$~1 is assumed here.}
\label{Fig:3}
\end{figure}

Fig.~\ref{Fig:3}(a)-(c) show the temperature dependence of $\kappa$ of a 1 $\mu$m long and 1 $\mu$m wide stanene sheet for the temperature range of 2~$<T<$~1000 K along the ZZ and AC directions with $p=$~1. The total lattice thermal conductivity ($\kappa_\text{total}=\sum\limits_{s} \kappa_s$) at low temperatures is found to initially increase with temperature, till it reaches a maximum, and subsequently decreases at higher temperatures (Fig.~\ref{Fig:3}(a)). $\kappa_\text{total}$ attains a peak value of 26.93 W/m-K at a temperature ($T_\text{max}$)~$\sim$ 34 K along ZZ and 16.78 W/m-K at $T_\text{max}\sim$ 40 K along AC (Fig.~\ref{Fig:3}(a)). The contribution to lattice thermal conductivity from optical phonons is negligible (not shown in Fig.~\ref{Fig:3}). With regard to the phonon scattering mechanisms involved in determining $\kappa_\text{total}$, the phonon-boundary scattering is found to be dominant for $T<$~30 K. However, at higher temperatures, the lattice thermal conductivity is determined by the phonon-phonon scatterings. We observe that in the temperature range of $T<$~410 K, $\kappa_\text{total}$ along ZZ is higher compared to AC and beyond 410 K, this trend reverses. This reversal is due to lower group velocities $(v_{g}(\omega_{s})_{j})$ along ZZ compared to AC and the dominance of phonon-boundary scattering ($(\tau_{sB})^{-1} \propto v_{g}(\omega_{s})_{j}$) at low temperatures, and phonon-phonon ($(\tau_{sU,N})^{-1} \propto (v_{g}(\omega_{s})_{j})^{-2}$) at high temperatures.

Next, we study the relative contributions from each phonon branch to $\kappa_\text{total}$ of stanene. The individual contributions from LA, TA, and ZA phonon branches along AC and ZZ directions are examined and are shown in Fig.~\ref{Fig:3}(b) and (c). At low temperatures ($T<$~10 K), the contribution to $\kappa_\text{total}$ originates predominantly from the ZA and TA branches, whereas at higher temperatures, the LA branch begins contributing dominantly. Due to large values of $\gamma_{ZA/TA}(\mathbf{q})$ compared to $\gamma_{LA}(\mathbf{q})$ (Fig.~\ref{Fig:2}(c)), the phonon-phonon scattering processes for the ZA and TA branches dominate at lower temperatures compared to the LA branch, leading to a suppression of ZA and TA thermal conductivity and a corresponding shift in their peaks to lower temperatures (Fig.~\ref{Fig:3}(b) and (c)). Furthermore, the contributions from all the branches along ZZ are greater than those along AC upto a temperature of 375 K. Beyond 375 K, the contribution from the LA branch along the AC direction begins to dominate over the corresponding branch along ZZ leading to a crossover in Fig.~\ref{Fig:3}(a) at 410 K. This is attributed to high group velocity and very low Gr\"{u}neisen parameter ($\kappa\propto \tau_\mathrm{U/N}\propto v_\mathrm{LA}^2/\gamma_\mathrm{LA}^2$) as seen in Fig.~\ref{Fig:2}(a, b, and c). From the applications perspective, it is important to study the thermal transport at room temperature (300 K). Unlike at low temperatures, where the contribution from the ZA branch is the maximum, the contributions at $T\geq$~300 K to $\kappa_\text{total}$ originate predominantly from the LA branch ($\sim$91\%), followed by the TA ($\sim$8\%) and ZA ($\sim$1\%) branches. The room temperature thermal conductivity of stanene along the ZZ and AC directions are 10.83 W/m-K and 9.2 W/m-K, respectively, indicating an anisotropy ($\sim$15\%) in the thermal transport. This is unlike graphene, which exhibits a nearly isotropic ($<$0.1\% anisotropy) thermal transport~\cite{kong2009first,jiang2009thermal}.

The lattice thermal conductivity of stanene is much lower than its bulk counterpart~\cite{lees1908effects,bartkowski1977anisotropy}, thick films of tin~\cite{kaul2014carrier}, silicene~\cite{gu2015first}, monolayer MoS$_{2}$~\cite{yan2014thermal,liu2014anisotropic}, germanene~\cite{balateroa2015molecular}, and phosphorene (blue and black)~\cite{liu2015anisotropic,jain2015strongly,zhang2016thermal}. As mentioned earlier, since the LA branch contributes the most to the thermal conductivity, we calculate Debye temperature ($\Theta_D$ $=\frac{\hbar\omega_\text{max}}{k_{B}}$) for the LA branch and find it to be around 72 K, about 16 times lower than that of graphene (1160 K). The main reasons for this low $\Theta_D$ of stanene are heavier atomic mass of tin and weaker bonding between tin atoms ($sp^\text{2}$-$sp^\text{3}$ hybridization) compared to graphene ($sp^\text{2}$ hybridization). This low Debye temperature leads to low lattice thermal conductivity of stanene compared to other elemental sheets belonging to the same group. 

To gain further insight into the nature of thermal transport in stanene, it is essential to evaluate the phonon mean free path (MFP). It gives an estimate of the average distance over which the phonons travel without losing energy and can provide an understanding of the thermal conduction processes and the scattering mechanisms operating in the system. Here, we study the temperature dependence of the effective phonon MFP $(\lambda_\text{eff})$ in stanene using~\cite{balandin2011thermal,liu2015anisotropic,liu2014anomalous}:
\begin{equation}
\lambda_\text{eff}(T)= \frac{\kappa_{j}(T)}{\sum\limits_{s}\int_{0}^{\omega_{max}} C_{v}(\omega_{s})\lbrace v_{g}(\omega_{s})\rbrace_{j}d\omega}
\label{eq:8}
\end{equation}  
  
where $C_{v} = \frac{\hbar^{2}}{2k_{B}T^{2}\delta}\bar{n}(\bar{n}+1)\omega_{s}^{2}g(\omega_{s})$ and $\kappa_{j}(T)$ is calculated using eqn~\ref{eq:3}. Fig.~\ref{Fig:3}(d) depicts the dependence of $\lambda_\text{eff}$ on $T$ along ZZ direction using the same parameters chosen for the calculation of lattice thermal conductivity. It is found that at very low temperatures, the contributions from LA and TA phonon branches are nearly constant (as shown by the almost flat regions in Fig.~\ref{Fig:3}(d)) which is characteristic of the pure phonon-boundary scattering. However, at higher temperatures, the MFPs decrease rapidly, due to the dominance of phonon-phonon scattering. The overall trend in the total $\lambda_\text{eff}$ is influenced by the ZA and TA branches at low temperatures ($T <$~20 K) and the LA branch at higher temperatures. The total $\lambda_\text{eff}$ at room temperature is around 17 nm which is about four times lower than phosphorene $(\lambda_\text{eff} \sim$~65.4 nm)~\cite{liu2015anisotropic}, nearly 21 times lower than bulk tin $(\lambda_\text{eff} \sim$~354 nm)~\cite{kaul2014carrier} and 45 times lower than graphene $(\lambda_\text{eff}\sim$~750 nm)~\cite{balandin2011thermal}. Such a low MFP in stanene indicates that the thermal transport is completely diffusive in nature.
 
\begin{figure}[!t]
\includegraphics[width=\columnwidth]{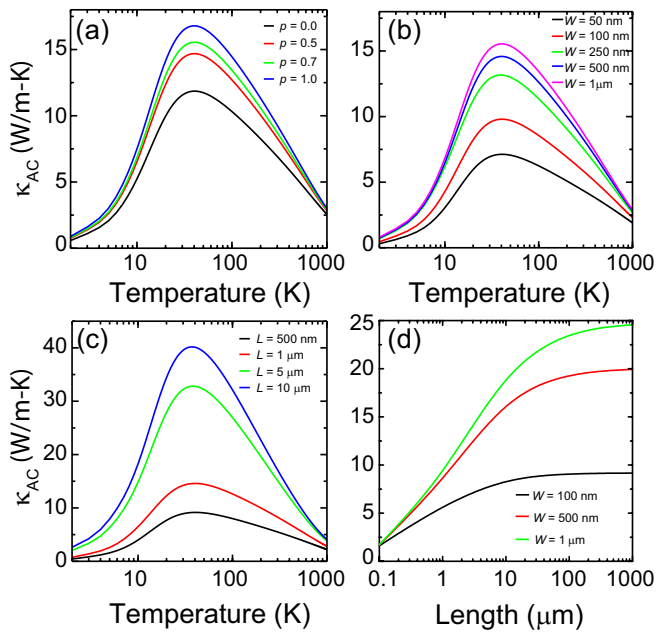}
\caption{Temperature variation of the total lattice thermal conductivity of stanene showing the influence of sample-dependent scattering parameters: (a) specular parameter $p$, (b) width ($W$), and (c) length ($L$) of the sample. (d) Variation of total lattice thermal conductivity with $L$ and $W$ of the stanene sample at room temperature ($T=$~300 K). The transport direction is along the AC direction.}
\label{Fig:4}
\end{figure} 

We next study the degree to which the lattice thermal conductivity of stanene along the AC direction ($\kappa_\text{AC}$) can be influenced by specularity ($p$) as well as width ($W$) and length ($L$) of the sample. The influence of edge roughness (through the parameter $p$) on $\kappa_\text{AC}$ for a 1 $\mu$m wide and 1 $\mu$m long stanene sample is shown in Fig.~\ref{Fig:4}(a). With the increase in the edge roughness (that is, decrease in $p$) of stanene, not only does the value of $\kappa_\text{AC}$ decreases, but also, its temperature dependence changes. In Fig.~\ref{Fig:4}(b) (Fig.~\ref{Fig:4}(c)), the variation in $\kappa_\text{AC}$ with the width (length) of the stanene sample is shown for $L =$~1 $\mu$m ($W=$~500 nm) and $p =$~0.7. Although the general features of $\kappa_\text{AC}$ are similar, the nature of the variation for each width (length) depends on the relative dominance of the individual contributions from LA, TA, and ZA branches. With the decrease in stanene width (length), over a wide range of temperatures, $\kappa_\text{AC}$ decreases and the corresponding $T_\text{max}$ shifts to higher values. The reduction of $\kappa_\text{AC}$ with decrease in length (width) is larger for temperatures below 100 K. This is because of the increase in the influence of phonon-boundary scattering and reduction in that of phonon-phonon scattering. Compared to other 2D materials such as graphene~\cite{nika2012anomalous,nissimagoudar2014significant} and MoS$_\text{2}$~\cite{li2013thermal}, the shift in the position of $\kappa_\text{AC}$ peak with temperature is very minute. Fig.~\ref{Fig:4}(d) shows the length dependence of $\kappa_\text{AC}$ corresponding to three different widths for a given specularity parameter, $p = $~0.7 and temperature, $T=$~300 K. $\kappa_\text{AC}$ increases more rapidly at lower values of $L$ and then tends to saturate at higher values. The rate of increase in $\kappa_\text{AC}$ with $L$ is more for higher values of $W$. At smaller sample sizes, the thermal conductivity is completely dominated by the phonon-boundary scattering upto a certain finite length ($\sim L=$~10 $\mu$m), beyond which, the intrinsic anharmonic phonon-phonon scattering becomes dominant. As a result, for a small sample size of 100$\times$100 nm$^{2}$ ($L\times W$), we obtain a low value of 1.6 W/m-K. Further reductions in the lattice thermal conductivity of stanene can be achieved by creating isotopic point defects/impurities and by reducing the dimensionality of the sample (e.g. stanene nanoribbons).

\section{Conclusions}
In summary, using the phonon Boltzmann transport formalism and full-band phonon dispersion relations obtained from first principles calculations, we predict the lattice thermal conductivity of stanene to be very low. This is because, stanene has very high Gr\"{u}neisen parameter and low group velocity, thereby indicating significantly high anharmonicity in this material. Unlike graphene, which exhibits nearly isotropic ($<$0.1\% anisotropy) thermal transport, stanene exhibits anisotropy ($\sim$15\%). The room temperature effective phonon mean free path is found to be nearly 45 times lower than graphene indicating that the thermal transport in stanene is completely diffusive in nature. At room temperature, the maximum contribution to the lattice thermal conductivity comes from the LA branch (91$\%$), unlike graphene, where the ZA branch is the main contributor. Furthermore, the lattice thermal conductivity can be further reduced by tuning the sample size and creating the rough surfaces at the edges as well as impurities/point defects. These findings show that the lattice thermal conductivity of stanene can be tuned to a large extent making it promising for thermoelectric applications.\\

\section*{Acknowledgements}
The authors acknowledge financial support from DST Nanomission. The authors thank the Supercomputing Education and Research Centre and Materials Research Centre, IISc, for the required computational facilities.

\bibliography{manuscript}

\end{document}